\documentclass[aps,twocolumn,pra,showpacs,10pt]{revtex4-1}

\usepackage{epsfig,amssymb,amsmath,graphicx,color}
\usepackage[dvips]{hyperref}

\begin{document}

\title{Scattering induced spatial superpositions in multi-particle localization}
\author{James S. Douglas and Keith Burnett}
\affiliation{University of Sheffield, Western Bank, Sheffield S10 2TN, United Kingdom}
\begin{abstract}
We describe how quasi-classical relative positions of particles emerge in an initially delocalized quantum system as scattering of a probe beam is observed.
We show that in the multi-particle case this localization in position space occurs via intermediate states that are quantum superpositions of spatial configurations. These superpositions are robust to consecutive scattering events and scattering alone does not lead to their decoherence. Instead the free evolution of the system combines with scattering to destroy these superpositions, leading the particles to adopt classical-like positions relative to one another. 
\end{abstract}
\maketitle

\section{Introduction}

We take for granted in our everyday experience that the distances between the objects around us are well defined and single valued in time. We can then formulate classical descriptions of how these objects and their relative positions behave when forces are applied to them. We know, however, that on the microscopic level systems follow the laws of quantum mechanics, laws that allow the relative positions of particles to exist in superpositions of different values. In this article we describe a new route from the quantum to the classical as the positions of particles are localized by consecutive scattering of probe particles. In particular we see that the transition happens via an intermediate phase that retains features of both the quantum and classical worlds.

The reduction of quantum superpositions to classical states is often explained using decoherence theory \cite{Zurek2003}, where interactions between the system being studied and its environment lead to superposition states of the system being converted to mixtures of classical-like states. A fundamental way a system can interact with its environment is to scatter particles from the environment, which for a single particle system reduces interference fringes at the output of an interferometer  \cite{Joos1985a,Gallis1990a,Pfau1994a,Chapman1995a,Hornberger2003a,Hornberger2003b,Kokorowski2001a,Uys2005a}. 
For a pair of particles, prepared in a delocalized quantum state, scattering from the system leads to localization of the particles relative to one another \cite{Rau2003a,Cable2005a,Zheng2005a,Dunningham2012a}, where each scattering event transfers information about the relative position to the environment. However, this mechanism by itself is not enough to localize systems of more than two particles.

Here we show, for a more general multi-particle case, how relative position is established by scattering probe particles and measuring the scattering angle. Each measurement of a scattering event leads to more information about the relative position of the system particles being transferred to the environment and the system many-body state localizes in relative position space. However, as we will see, the measured scattering pattern does not distinguish between two spatially distinct configurations and the system forms a coherent spatial superposition. These superpositions form when scattering is fast compared with free evolution of the many-body system and they are robust against further scattering and could be detected. On longer time scales we show that the superpositions are eventually destroyed through a combination of scattering and momentum evolution, thus establishing a model of how the classical-like behavior we are familiar with originates from the underlying quantum mechanics.

\section{Scattering from multiple particles}

\begin{figure}
\centering
\includegraphics{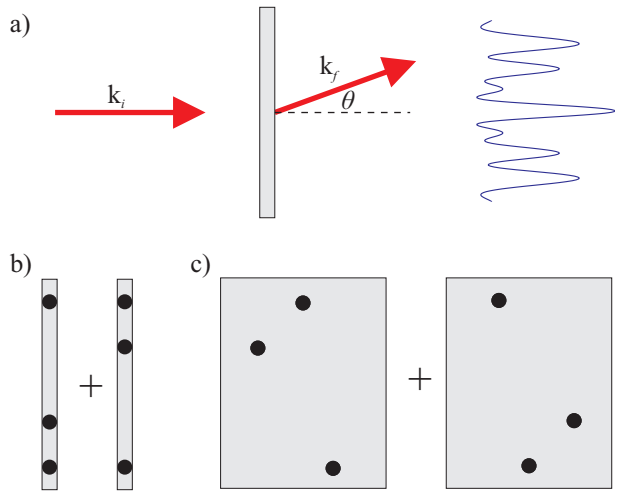}
\caption{(a) Scattering from the system of particles leads to a change of probe wave-vector from $\mathbf{k}_i$ to $\mathbf{k}_f$ with associated scattering angle $\theta$. A typical scattering probability distribution is shown on the right. (b) For a one dimensional system of particles, scattering leads to superpositions of particle positions of the type shown, where the two configurations both have the same set of interparticle distances. (c) Similarly, scattering from a two dimensional system leads to superpositions of configurations where the positions are rotated by $180^{\circ}$.}
\label{fig:scatter_superposition}
\end{figure}

We examine the effect on the many-body state of consecutive coherent scattering and detection events by looking at the scattering problem from a general perspective that allows us to ignore the details of the interaction potential involved. We consider a system of identical particles with associated field creation and annihilation operators $\hat{\Psi}^\dagger(\mathbf{r})$ and $\hat{\Psi}(\mathbf{r})$. The probe particles are assumed to be approximately in plane wave form with associated wave-vector $\mathbf{k}_i$. Coherent scattering of the probe by the system is then associated with a momentum transfer to the system of $\hbar\mathbf{k} =\hbar(\mathbf{k}_i - \mathbf{k}_f)$, where $\mathbf{k}_f$ is the probe wave-vector after scattering, as shown in Fig.~\ref{fig:scatter_superposition}(a).
Measurement of the final probe wave-vector, for example by imaging in the far field, then transforms the system state according to
\begin{equation}
|\psi\rangle \longrightarrow \int d\mathbf{r} \hat{\Psi}^\dagger(\mathbf{r})\hat{\Psi}(\mathbf{r})e^{i\mathbf{k}\cdot \mathbf{r}}|\psi\rangle.
\label{eq:scatter_transform}
\end{equation}
For an initial $N$-particle state $|\psi\rangle = \int d\mathbf{R} \phi(\mathbf{R})|\mathbf{R}\rangle$, where  $\mathbf{R}= \{\mathbf{r}_1,\ldots,\mathbf{r}_N\}$ gives the coordinates of the $N$ particles, the associated probability of observing a scattered probe wave-vector of $\mathbf{k}_f$ is
\begin{equation}
P(\mathbf{k}_f) = \frac{g^2}{4 \pi}\int d\mathbf{R}\left|\phi(\mathbf{R})\sum_{j=1}^N e^{i \mathbf{k}\cdot\mathbf{r}_j}\right|^2.
\label{eq:prob_scatter}
\end{equation}
The factor $g$ is determined by the strength of the probe-system interaction and is assumed to be small so that multiple scattering of an individual probe particle is negligible. 
For simplicity we assume $g$ is independent of $\mathbf{k}_f$, which is the case when energy conservation restricts scattering so that $|\mathbf{k}_f|\sim|\mathbf{k}_i|$ and the probe-system interaction is isotropic \cite{VanHove1954}.
This type of scattering can be realized in systems of ultracold atoms where isotropic $s$-wave scattering dominates other scattering processes \cite{Sanders2010a}. Anisotropic scattering, such as the dipole pattern in light scattering, can be taken into account by straightforward generalization.


The transformation of the state in Eq.~(\ref{eq:scatter_transform}) projects the system's state toward configurations with higher probability of scattering in the measured direction.
The probability of scattering increases as the particles' relative positions become localized, and observing a scattering event projects the state towards greater localization of the particles. 
The scattering distribution depends on $\left|\sum_{j=1}^N e^{i \mathbf{k}\cdot\mathbf{r}_j}\right|^2$, which is completely determined by the relative positions of pairs of particles. The information gained from a measurement of the scattering distribution is therefore limited to relative position and the measurement will then preserve superpositions of configurations that have the same set of relative position vectors. Examples of this type of superposition are shown in Fig.~\ref{fig:scatter_superposition}(b) and (c) for a three particle system. Each configuration in such a superposition will give the same scattering pattern and consecutive scattering measurements can push the system towards this type of superposition.
A similar effect is observed in arrays of Bose-Einstein condensates, where measurement of atoms coupled out of the condensates results in superposition of relative phase \cite{Dunningham2006a}.

Apart from scattering we must also consider the situation when the probe does not scatter from the system. Rather than leaving the system in the same state, the measurement of a non-scattering event also leads to a change in the state \cite{Rau2003a}. This is because non-scattering is more likely from certain particle distributions and detecting a non-scattering event projects the wavefunction of the system towards these configurations. For a position eigenstate $|\mathbf{R}\rangle$ the probability of non-scattering is determined by the conservation of probe particles, that is the probe particles must either be scattered or not scattered, giving 
\begin{equation}
P_{NS}(\mathbf{R}) = A(\mathbf{R})^2= 1 -  \frac{g^2}{4 \pi}\int d \Omega_{\mathbf{k}_f} \left|\sum_{j=1}^N e^{i \mathbf{k}\cdot\mathbf{r}_j}\right|^2,
\end{equation}
where the integration is over all scattering angles.
Detecting a non-scattering event then transforms the system state according to 
\begin{equation}
|\psi\rangle \longrightarrow \int d\mathbf{R} A(\mathbf{R})\phi(\mathbf{R}) |\mathbf{R}\rangle,
\label{eq:non_scatter_transform}
\end{equation}
and occurs with total probability $\int d\mathbf{R} |A(\mathbf{R})\phi(\mathbf{R})|^2$.

Consecutive scattering and non-scattering events lead to a dynamic evolution of the many-body state. The dynamic scattering process can be simulated by the following quantum jump procedure \cite{Dalibard1992a,Dum4879a}. Taking the initial state, we calculate the probability distribution for scattering and non-scattering. Using this distribution we randomly ``detect'' an event, and apply either the projection Eq.~(\ref{eq:non_scatter_transform}) following detection of non-scattering or the projection Eq.~(\ref{eq:scatter_transform}) for scattering detected at a particular angle. In either case the many-body state is then normalized and becomes the input state and the process repeats. In between each scattering event we evolve the system according to the free space Hamiltonian $H_0 = \sum_j p_j^2/2m$. For simplicity we assume the scattering events are equally spaced in time and the system evolves for time $dt$ between each event.


\begin{figure}
\centering
\includegraphics{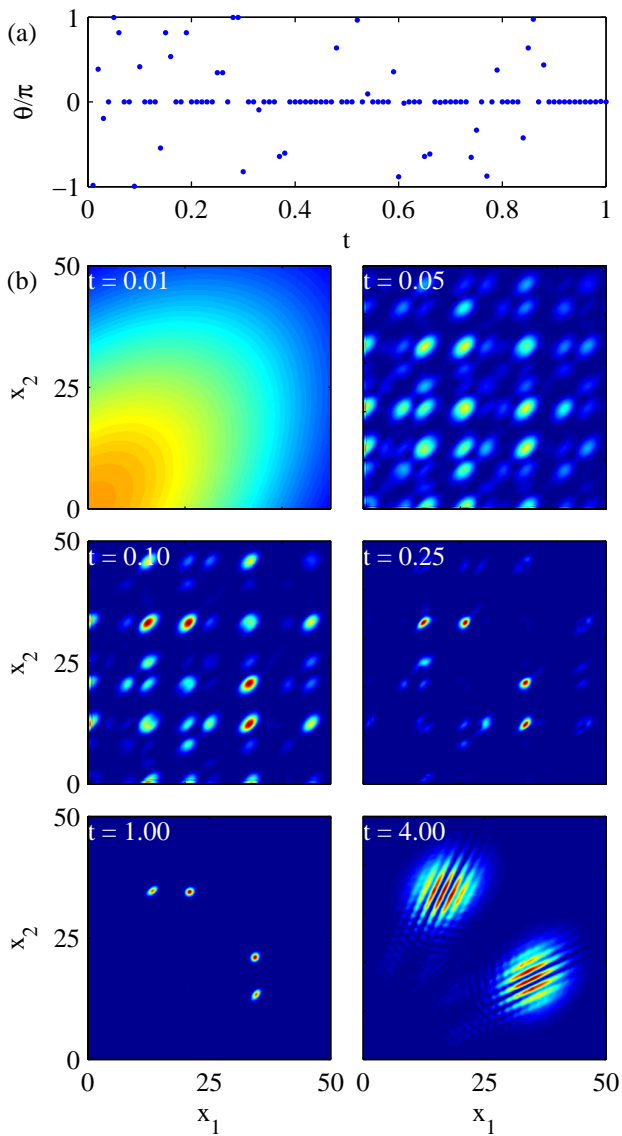}
\caption{Example evolution of a three particle system from a delocalized state into a superposition of localized position states. (a) Scattering events plotted over time. (b) Evolution of the system's probability density in relative position space. The scattering interaction is turned on at $t=0$ until $t=1$ and the state is then allowed to expand for until $t=4$, shown in the last frame.}
\label{fig:develop_three_site}
\end{figure}


\section{Three particle localization}

We now consider an example calculation for a three particle system. To make the simulations tractable we assume the particles are confined to one dimension \cite{[{Scattering from an effectively one-dimensional gas of cold atoms was recently demonstrated in }] Gadway2011a} and that the scattering of probe particles is confined to a two dimensional plane with angle $\theta$ as shown in Fig.~\ref{fig:scatter_superposition}(a). We assume that the initial wavefunction $\phi(r_1,r_2,r_3)$ can be decomposed into $\Phi(X)\phi(x_1,x_2)$ where $X = (r_1+r_2+r_3)/3$ is the center of mass and $x_1 = r_2-r_1$ and $x_2 = r_3 - r_1$ are relative coordinates with units $1/k_i$. As the scattering operators affect the center of mass and the relative positions independently, the system remains separable after scattering. The wavefunction can be expressed similarly in wave vector space as $\Phi(K)\phi(q_1,q_2)$ where $K = (k_1+k_2+k_3)/3$ is the average wave vector of the particles and $q_1 =(2k_2-k_1-k_3)/3$ and $q_2 = (2k_3-k_1-k_2)/3$ are relative wave vectors. The evolution due to $H_0$ then also remains separable with $\Phi(K,t+dt) = e^{-3 i K^2 dt}\Phi(K,t)$ and $\Phi(q_1,q_2,t+dt) = e^{-2 i (q_1^2+q_2^2+q_1 q_2) dt}\Phi(q_1,q_2,t)$, where our times are expressed in units of $2 m/(\hbar k_i^2)$. 
The center of mass then evolves due to the momentum added or subtracted from the entire system by the scattering events, while the relative coordinates evolve independently. We note that the separability of the wavefunction is not crucial to our results, and indeed we observe the same behavior in simulations with non-separable wavefunctions in the full three dimensional configuration space.

In Fig.~\ref{fig:develop_three_site} we show an example evolution for the three particle system starting from an initially delocalized state with $\phi(x_1,x_2)$ equal to a constant over the simulation region. Fig.~\ref{fig:develop_three_site}(a) shows the detection events that occur, the majority of which are non-scattering events shown as $\theta = 0$. The resulting localization is shown in Fig.~\ref{fig:develop_three_site}(b) and occurs quickly with only 100 events leading to the well localized state at $t=1$. Note that the probability density is symmetric in exchange of $x_1$ and $x_2$ as required for identical particles. Despite the localization, at $t=1$ the system maintains a superposition of the particles having $x_1\sim 14$ and $x_2\sim 35$ and having $x_1\sim 21$ and $x_2\sim 35$. This is an example of the type of superposition shown in Fig.~\ref{fig:scatter_superposition}(b) and forms an intermediate step in the transition between the delocalized quantum state and a quasi-classical state with definite relative position. This superposition can be detected by allowing the particle wavefunction to expand under free evolution by turning off the scattering interaction, where after a suitable period of time the wavefunction shows interference fringes, as shown in the last frame of Fig.~\ref{fig:develop_three_site}(b). Although the phase of this interference varies between each realization, the Bayesian analysis used in Ref.~\cite{Dunningham2012a} could be used here to detect the presence of the scattering induced superpositions. For more than three particles superpositions would not necessarily produce simple interference patterns but would be revealed in higher order correlation functions \cite{Altman2004a}.


\begin{figure}
\centering
\includegraphics{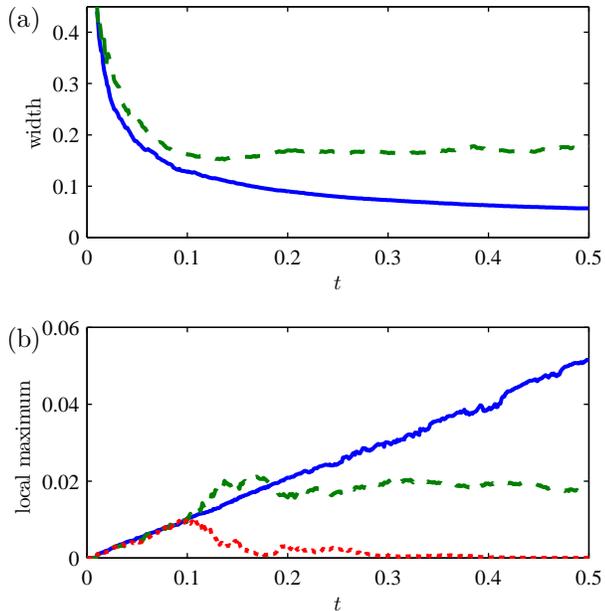}
\caption{Comparison of localization during scattering with and without $H_0$-evolution of the wavefunction between scattering events. (a) Width of one of the localized peaks in the probability density as a function of time without $H_0$-evolution (solid) and with $H_0$-evolution (dashed). (b) Change in value of the probability density at the two localized peaks of the superposition. Without $H_0$-evolution the two peaks have equal value (solid), while with $H_0$-evolution one peak (dashed) begins to dominate over the other (dotted) after an initial period.}
\label{fig:Localisation_and_momentum}
\end{figure}


Localization of the particles happens due to the uncertainty about which particle scattered the probe. When a scattering event is detected it implies the system is then in a superposition of each of the particles having received momentum kick $\mathbf{k}$. This widens the relative momentum distribution leading to a corresponding narrowing of the spatial distribution. By the central limit theorem, after $N$ random scattering events we expect the momentum distribution's width to scale as $\sqrt{N}{k}_i$ and consequently expect the width of the spatial distribution to scale as $1/(\sqrt{N}{k}_i)$. However, this ignores the expansion of the particles' wavefunction due to free evolution under $H_0$, which we expect will counter the localization process \cite{Kiefer2000a}. This behavior is shown in Fig.~\ref{fig:Localisation_and_momentum}(a) for a simulation of the three particle localization resulting from 5000 detection events with time spacing $dt=0.0001$. Here we compare the case where there is $H_0$-evolution between scattering events and the case where free evolution is artificially turned off (equivalent to $dt=0$). Without $H_0$ evolution the particles localize relative to one another with the expected $1/\sqrt{N}$ scaling. When the free evolution is included, the localization is initially similar, but then the free expansion begins to oppose the localization pressure from scattering and the width of the wavefunction plateaus.

\begin{figure}
\centering
\includegraphics{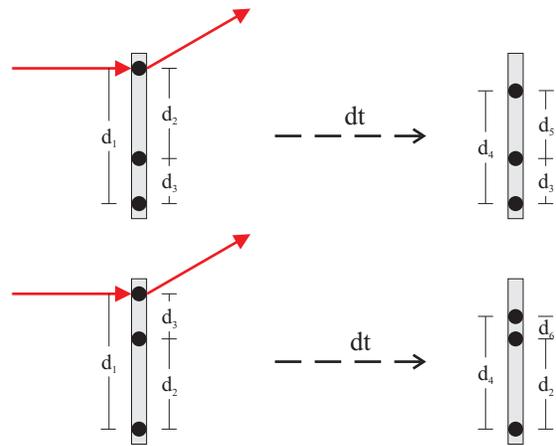}
\caption{Relative position change caused by $H_0$ evolution after one of the particles scatters a probe particle. The top and bottom configurations initially have indistinguishable scattering patterns because they share the same set of interparticle distances $\{d_1,d_2,d_3\}$. After one particle scatters the probe its position changes in time and the configurations become distinguishable as they acquire different sets of interparticle distances $\{d_3,d_4,d_5\}\neq\{d_2,d_4,d_6\}$.}
\label{fig:Relative_position_change}
\end{figure}

The free space evolution also affects the superposition that results from the initial localization. In Fig.~\ref{fig:Localisation_and_momentum}(b) we show the evolution of the local maxima of the peaks in the wavefunction associated with the two configurations in the spatial superposition. Without $H_0$ evolution the two configurations maintain equal weight throughout the scattering process, while including free evolution results in one of the configurations eventually being preferred and the superposition diminishes. This occurs because the two different particle configurations in the superposition evolve freely to have different relative positions and scattering then projects the state so one of the configurations dominates. This process is shown in Fig.~\ref{fig:Relative_position_change} where two configurations with identical scattering patterns evolve into configurations with different scattering patterns, whereby continued scattering will lead to decoherence of the superposition. Thus the localized quantum superpositions we saw in Fig.~\ref{fig:develop_three_site} only exist for the intermediate time between the start of scattering and the natural free evolution time of the system. For longer time periods the superpositions are destroyed by scattering combined with free evolution and quasiclassical relative positions are established between the particles.


\section{Discussion and conclusion}

The scattering interaction we have described above provides a partial measurement of the relative positions of particles in a system. Each scattering event then leads to the system's environment containing more information about the position state and in the limit of an infinite number of events the measurement is projective onto perfectly localized states $|\mathbf{R}\rangle$.
The localized states have uniform momentum distributions and the momentum shift caused by scattering leaves these states unchanged. The environment can then gain increasing amounts of information about the localized states through the scattering interaction without disturbing them and these states become the basis for classical-like reality. In the language of quantum Darwinism these states are the ``fittest'' because multiple copies of the information about the state can build up in the environment without altering the state \cite{Zurek2004a,Zurek2006a,Zurek2007a,Zurek2010a}. 

The interesting consequence of this particular interaction is that distinctly non-classical superpositions of the localized states are preserved by the scattering process. This gives a twist to the simplest version of quantum Darwinism, which suggests the build up of redundant information in the environment should only support states of the systems that behave classically.
Rather, we see that it is only when the scattering interaction happens on a time scale that is comparable with the free evolution of the system that the superposition begins to decohere. Our model then shows how, as the superposition is destroyed, the interparticle distances take on the robust, single values that we are familiar with. In doing so we have highlighted that the interaction with the environment alone is not enough to ensure classicality but that the system's free evolution also plays an important role. 

In conclusion, we have shown how observation of a system of particles, while eventually leading to a quasi-classical description, can create and preserve spatial superposition states. This is a significant observation that needs to be folded into the overall story of decoherence and the quantum to classical transition.



\begin{thebibliography}{25}%
\makeatletter
\providecommand \@ifxundefined [1]{%
 \@ifx{#1\undefined}
}%
\providecommand \@ifnum [1]{%
 \ifnum #1\expandafter \@firstoftwo
 \else \expandafter \@secondoftwo
 \fi
}%
\providecommand \@ifx [1]{%
 \ifx #1\expandafter \@firstoftwo
 \else \expandafter \@secondoftwo
 \fi
}%
\providecommand \natexlab [1]{#1}%
\providecommand \enquote  [1]{``#1''}%
\providecommand \bibnamefont  [1]{#1}%
\providecommand \bibfnamefont [1]{#1}%
\providecommand \citenamefont [1]{#1}%
\providecommand \href@noop [0]{\@secondoftwo}%
\providecommand \href [0]{\begingroup \@sanitize@url \@href}%
\providecommand \@href[1]{\@@startlink{#1}\@@href}%
\providecommand \@@href[1]{\endgroup#1\@@endlink}%
\providecommand \@sanitize@url [0]{\catcode `\\12\catcode `\$12\catcode
  `\&12\catcode `\#12\catcode `\^12\catcode `\_12\catcode `\%12\relax}%
\providecommand \@@startlink[1]{}%
\providecommand \@@endlink[0]{}%
\providecommand \url  [0]{\begingroup\@sanitize@url \@url }%
\providecommand \@url [1]{\endgroup\@href {#1}{\urlprefix }}%
\providecommand \urlprefix  [0]{URL }%
\providecommand \Eprint [0]{\href }%
\providecommand \doibase [0]{http://dx.doi.org/}%
\providecommand \selectlanguage [0]{\@gobble}%
\providecommand \bibinfo  [0]{\@secondoftwo}%
\providecommand \bibfield  [0]{\@secondoftwo}%
\providecommand \translation [1]{[#1]}%
\providecommand \BibitemOpen [0]{}%
\providecommand \bibitemStop [0]{}%
\providecommand \bibitemNoStop [0]{.\EOS\space}%
\providecommand \EOS [0]{\spacefactor3000\relax}%
\providecommand \BibitemShut  [1]{\csname bibitem#1\endcsname}%
\let\auto@bib@innerbib\@empty
\bibitem [{\citenamefont {Zurek}(2003)}]{Zurek2003}%
  \BibitemOpen
  \bibfield  {author} {\bibinfo {author} {\bibfnamefont {W.~H.}\ \bibnamefont
  {Zurek}},\ }\href {\doibase 10.1103/RevModPhys.75.715} {\bibfield  {journal}
  {\bibinfo  {journal} {Rev. Mod. Phys.}\ }\textbf {\bibinfo {volume} {75}},\
  \bibinfo {pages} {715} (\bibinfo {year} {2003})}\BibitemShut {NoStop}%
\bibitem [{\citenamefont {Joos}\ and\ \citenamefont {Zeh}(1985)}]{Joos1985a}%
  \BibitemOpen
  \bibfield  {author} {\bibinfo {author} {\bibfnamefont {E.}~\bibnamefont
  {Joos}}\ and\ \bibinfo {author} {\bibfnamefont {H.~D.}\ \bibnamefont {Zeh}},\
  }\href {http://dx.doi.org/10.1007/BF01725541} {\bibfield  {journal} {\bibinfo
   {journal} {Zeitschrift für Physik B Condensed Matter}\ }\textbf {\bibinfo
  {volume} {59}},\ \bibinfo {pages} {223} (\bibinfo {year} {1985})},\ \bibinfo
  {note} {10.1007/BF01725541}\BibitemShut {NoStop}%
\bibitem [{\citenamefont {Gallis}\ and\ \citenamefont
  {Fleming}(1990)}]{Gallis1990a}%
  \BibitemOpen
  \bibfield  {author} {\bibinfo {author} {\bibfnamefont {M.~R.}\ \bibnamefont
  {Gallis}}\ and\ \bibinfo {author} {\bibfnamefont {G.~N.}\ \bibnamefont
  {Fleming}},\ }\href {\doibase 10.1103/PhysRevA.42.38} {\bibfield  {journal}
  {\bibinfo  {journal} {Phys. Rev. A}\ }\textbf {\bibinfo {volume} {42}},\
  \bibinfo {pages} {38} (\bibinfo {year} {1990})}\BibitemShut {NoStop}%
\bibitem [{\citenamefont {Pfau}\ \emph {et~al.}(1994)\citenamefont {Pfau},
  \citenamefont {Sp\"alter}, \citenamefont {Kurtsiefer}, \citenamefont
  {Ekstrom},\ and\ \citenamefont {Mlynek}}]{Pfau1994a}%
  \BibitemOpen
  \bibfield  {author} {\bibinfo {author} {\bibfnamefont {T.}~\bibnamefont
  {Pfau}}, \bibinfo {author} {\bibfnamefont {S.}~\bibnamefont {Sp\"alter}},
  \bibinfo {author} {\bibfnamefont {C.}~\bibnamefont {Kurtsiefer}}, \bibinfo
  {author} {\bibfnamefont {C.~R.}\ \bibnamefont {Ekstrom}}, \ and\ \bibinfo
  {author} {\bibfnamefont {J.}~\bibnamefont {Mlynek}},\ }\href {\doibase
  10.1103/PhysRevLett.73.1223} {\bibfield  {journal} {\bibinfo  {journal}
  {Phys. Rev. Lett.}\ }\textbf {\bibinfo {volume} {73}},\ \bibinfo {pages}
  {1223} (\bibinfo {year} {1994})}\BibitemShut {NoStop}%
\bibitem [{\citenamefont {Chapman}\ \emph {et~al.}(1995)\citenamefont
  {Chapman}, \citenamefont {Hammond}, \citenamefont {Lenef}, \citenamefont
  {Schmiedmayer}, \citenamefont {Rubenstein}, \citenamefont {Smith},\ and\
  \citenamefont {Pritchard}}]{Chapman1995a}%
  \BibitemOpen
  \bibfield  {author} {\bibinfo {author} {\bibfnamefont {M.~S.}\ \bibnamefont
  {Chapman}}, \bibinfo {author} {\bibfnamefont {T.~D.}\ \bibnamefont
  {Hammond}}, \bibinfo {author} {\bibfnamefont {A.}~\bibnamefont {Lenef}},
  \bibinfo {author} {\bibfnamefont {J.}~\bibnamefont {Schmiedmayer}}, \bibinfo
  {author} {\bibfnamefont {R.~A.}\ \bibnamefont {Rubenstein}}, \bibinfo
  {author} {\bibfnamefont {E.}~\bibnamefont {Smith}}, \ and\ \bibinfo {author}
  {\bibfnamefont {D.~E.}\ \bibnamefont {Pritchard}},\ }\href {\doibase
  10.1103/PhysRevLett.75.3783} {\bibfield  {journal} {\bibinfo  {journal}
  {Phys. Rev. Lett.}\ }\textbf {\bibinfo {volume} {75}},\ \bibinfo {pages}
  {3783} (\bibinfo {year} {1995})}\BibitemShut {NoStop}%
\bibitem [{\citenamefont {Hornberger}\ and\ \citenamefont
  {Sipe}(2003)}]{Hornberger2003a}%
  \BibitemOpen
  \bibfield  {author} {\bibinfo {author} {\bibfnamefont {K.}~\bibnamefont
  {Hornberger}}\ and\ \bibinfo {author} {\bibfnamefont {J.~E.}\ \bibnamefont
  {Sipe}},\ }\href {\doibase 10.1103/PhysRevA.68.012105} {\bibfield  {journal}
  {\bibinfo  {journal} {Phys. Rev. A}\ }\textbf {\bibinfo {volume} {68}},\
  \bibinfo {pages} {012105} (\bibinfo {year} {2003})}\BibitemShut {NoStop}%
\bibitem [{\citenamefont {Hornberger}\ \emph {et~al.}(2003)\citenamefont
  {Hornberger}, \citenamefont {Uttenthaler}, \citenamefont {Brezger},
  \citenamefont {Hackerm\"uller}, \citenamefont {Arndt},\ and\ \citenamefont
  {Zeilinger}}]{Hornberger2003b}%
  \BibitemOpen
  \bibfield  {author} {\bibinfo {author} {\bibfnamefont {K.}~\bibnamefont
  {Hornberger}}, \bibinfo {author} {\bibfnamefont {S.}~\bibnamefont
  {Uttenthaler}}, \bibinfo {author} {\bibfnamefont {B.}~\bibnamefont
  {Brezger}}, \bibinfo {author} {\bibfnamefont {L.}~\bibnamefont
  {Hackerm\"uller}}, \bibinfo {author} {\bibfnamefont {M.}~\bibnamefont
  {Arndt}}, \ and\ \bibinfo {author} {\bibfnamefont {A.}~\bibnamefont
  {Zeilinger}},\ }\href {\doibase 10.1103/PhysRevLett.90.160401} {\bibfield
  {journal} {\bibinfo  {journal} {Phys. Rev. Lett.}\ }\textbf {\bibinfo
  {volume} {90}},\ \bibinfo {pages} {160401} (\bibinfo {year}
  {2003})}\BibitemShut {NoStop}%
\bibitem [{\citenamefont {Kokorowski}\ \emph {et~al.}(2001)\citenamefont
  {Kokorowski}, \citenamefont {Cronin}, \citenamefont {Roberts},\ and\
  \citenamefont {Pritchard}}]{Kokorowski2001a}%
  \BibitemOpen
  \bibfield  {author} {\bibinfo {author} {\bibfnamefont {D.~A.}\ \bibnamefont
  {Kokorowski}}, \bibinfo {author} {\bibfnamefont {A.~D.}\ \bibnamefont
  {Cronin}}, \bibinfo {author} {\bibfnamefont {T.~D.}\ \bibnamefont {Roberts}},
  \ and\ \bibinfo {author} {\bibfnamefont {D.~E.}\ \bibnamefont {Pritchard}},\
  }\href {\doibase 10.1103/PhysRevLett.86.2191} {\bibfield  {journal} {\bibinfo
   {journal} {Phys. Rev. Lett.}\ }\textbf {\bibinfo {volume} {86}},\ \bibinfo
  {pages} {2191} (\bibinfo {year} {2001})}\BibitemShut {NoStop}%
\bibitem [{\citenamefont {Uys}\ \emph {et~al.}(2005)\citenamefont {Uys},
  \citenamefont {Perreault},\ and\ \citenamefont {Cronin}}]{Uys2005a}%
  \BibitemOpen
  \bibfield  {author} {\bibinfo {author} {\bibfnamefont {H.}~\bibnamefont
  {Uys}}, \bibinfo {author} {\bibfnamefont {J.~D.}\ \bibnamefont {Perreault}},
  \ and\ \bibinfo {author} {\bibfnamefont {A.~D.}\ \bibnamefont {Cronin}},\
  }\href {\doibase 10.1103/PhysRevLett.95.150403} {\bibfield  {journal}
  {\bibinfo  {journal} {Phys. Rev. Lett.}\ }\textbf {\bibinfo {volume} {95}},\
  \bibinfo {pages} {150403} (\bibinfo {year} {2005})}\BibitemShut {NoStop}%
\bibitem [{\citenamefont {Rau}\ \emph {et~al.}(2003)\citenamefont {Rau},
  \citenamefont {Dunningham},\ and\ \citenamefont {Burnett}}]{Rau2003a}%
  \BibitemOpen
  \bibfield  {author} {\bibinfo {author} {\bibfnamefont {A.~V.}\ \bibnamefont
  {Rau}}, \bibinfo {author} {\bibfnamefont {J.~A.}\ \bibnamefont {Dunningham}},
  \ and\ \bibinfo {author} {\bibfnamefont {K.}~\bibnamefont {Burnett}},\ }\href
  {\doibase 10.1126/science.1084867} {\bibfield  {journal} {\bibinfo  {journal}
  {Science}\ }\textbf {\bibinfo {volume} {301}},\ \bibinfo {pages} {1081}
  (\bibinfo {year} {2003})}\BibitemShut {NoStop}%
\bibitem [{\citenamefont {Cable}\ \emph {et~al.}(2005)\citenamefont {Cable},
  \citenamefont {Knight},\ and\ \citenamefont {Rudolph}}]{Cable2005a}%
  \BibitemOpen
  \bibfield  {author} {\bibinfo {author} {\bibfnamefont {H.}~\bibnamefont
  {Cable}}, \bibinfo {author} {\bibfnamefont {P.~L.}\ \bibnamefont {Knight}}, \
  and\ \bibinfo {author} {\bibfnamefont {T.}~\bibnamefont {Rudolph}},\ }\href
  {\doibase 10.1103/PhysRevA.71.042107} {\bibfield  {journal} {\bibinfo
  {journal} {Phys. Rev. A}\ }\textbf {\bibinfo {volume} {71}},\ \bibinfo
  {pages} {042107} (\bibinfo {year} {2005})}\BibitemShut {NoStop}%
\bibitem [{\citenamefont {Zheng}\ \emph {et~al.}(2005)\citenamefont {Zheng},
  \citenamefont {Li}, \citenamefont {Li},\ and\ \citenamefont
  {Sun}}]{Zheng2005a}%
  \BibitemOpen
  \bibfield  {author} {\bibinfo {author} {\bibfnamefont {L.}~\bibnamefont
  {Zheng}}, \bibinfo {author} {\bibfnamefont {C.}~\bibnamefont {Li}}, \bibinfo
  {author} {\bibfnamefont {Y.}~\bibnamefont {Li}}, \ and\ \bibinfo {author}
  {\bibfnamefont {C.~P.}\ \bibnamefont {Sun}},\ }\href {\doibase
  10.1103/PhysRevA.71.062101} {\bibfield  {journal} {\bibinfo  {journal} {Phys.
  Rev. A}\ }\textbf {\bibinfo {volume} {71}},\ \bibinfo {pages} {062101}
  (\bibinfo {year} {2005})}\BibitemShut {NoStop}%
\bibitem [{\citenamefont {{Knott}}\ \emph {et~al.}(2012)\citenamefont
  {{Knott}}, \citenamefont {{Sindt}},\ and\ \citenamefont
  {{Dunningham}}}]{Dunningham2012a}%
  \BibitemOpen
  \bibfield  {author} {\bibinfo {author} {\bibfnamefont {P.~A.}\ \bibnamefont
  {{Knott}}}, \bibinfo {author} {\bibfnamefont {J.}~\bibnamefont {{Sindt}}}, \
  and\ \bibinfo {author} {\bibfnamefont {J.~A.}\ \bibnamefont {{Dunningham}}},\
  }\href@noop {} {\bibfield  {journal} {\bibinfo  {journal} {ArXiv e-prints}\ }
  (\bibinfo {year} {2012})},\ \Eprint {http://arxiv.org/abs/1211.3555}
  {arXiv:1211.3555 [quant-ph]} \BibitemShut {NoStop}%
\bibitem [{\citenamefont {Van~Hove}(1954)}]{VanHove1954}%
  \BibitemOpen
  \bibfield  {author} {\bibinfo {author} {\bibfnamefont {L.}~\bibnamefont
  {Van~Hove}},\ }\href {\doibase 10.1103/PhysRev.95.249} {\bibfield  {journal}
  {\bibinfo  {journal} {Phys. Rev.}\ }\textbf {\bibinfo {volume} {95}},\
  \bibinfo {pages} {249} (\bibinfo {year} {1954})}\BibitemShut {NoStop}%
\bibitem [{\citenamefont {Sanders}\ \emph {et~al.}(2010)\citenamefont
  {Sanders}, \citenamefont {Mintert},\ and\ \citenamefont
  {Heller}}]{Sanders2010a}%
  \BibitemOpen
  \bibfield  {author} {\bibinfo {author} {\bibfnamefont {S.~N.}\ \bibnamefont
  {Sanders}}, \bibinfo {author} {\bibfnamefont {F.}~\bibnamefont {Mintert}}, \
  and\ \bibinfo {author} {\bibfnamefont {E.~J.}\ \bibnamefont {Heller}},\
  }\href {\doibase 10.1103/PhysRevLett.105.035301} {\bibfield  {journal}
  {\bibinfo  {journal} {Phys. Rev. Lett.}\ }\textbf {\bibinfo {volume} {105}},\
  \bibinfo {pages} {035301} (\bibinfo {year} {2010})}\BibitemShut {NoStop}%
\bibitem [{\citenamefont {Dunningham}\ \emph {et~al.}(2006)\citenamefont
  {Dunningham}, \citenamefont {Burnett}, \citenamefont {Roth},\ and\
  \citenamefont {Phillips}}]{Dunningham2006a}%
  \BibitemOpen
  \bibfield  {author} {\bibinfo {author} {\bibfnamefont {J.~A.}\ \bibnamefont
  {Dunningham}}, \bibinfo {author} {\bibfnamefont {K.}~\bibnamefont {Burnett}},
  \bibinfo {author} {\bibfnamefont {R.}~\bibnamefont {Roth}}, \ and\ \bibinfo
  {author} {\bibfnamefont {W.~D.}\ \bibnamefont {Phillips}},\ }\href
  {http://stacks.iop.org/1367-2630/8/i=9/a=182} {\bibfield  {journal} {\bibinfo
   {journal} {New Journal of Physics}\ }\textbf {\bibinfo {volume} {8}},\
  \bibinfo {pages} {182} (\bibinfo {year} {2006})}\BibitemShut {NoStop}%
\bibitem [{\citenamefont {Dalibard}\ \emph {et~al.}(1992)\citenamefont
  {Dalibard}, \citenamefont {Castin},\ and\ \citenamefont
  {M\o{}lmer}}]{Dalibard1992a}%
  \BibitemOpen
  \bibfield  {author} {\bibinfo {author} {\bibfnamefont {J.}~\bibnamefont
  {Dalibard}}, \bibinfo {author} {\bibfnamefont {Y.}~\bibnamefont {Castin}}, \
  and\ \bibinfo {author} {\bibfnamefont {K.}~\bibnamefont {M\o{}lmer}},\ }\href
  {\doibase 10.1103/PhysRevLett.68.580} {\bibfield  {journal} {\bibinfo
  {journal} {Phys. Rev. Lett.}\ }\textbf {\bibinfo {volume} {68}},\ \bibinfo
  {pages} {580} (\bibinfo {year} {1992})}\BibitemShut {NoStop}%
\bibitem [{\citenamefont {Dum}\ \emph {et~al.}(1992)\citenamefont {Dum},
  \citenamefont {Zoller},\ and\ \citenamefont {Ritsch}}]{Dum4879a}%
  \BibitemOpen
  \bibfield  {author} {\bibinfo {author} {\bibfnamefont {R.}~\bibnamefont
  {Dum}}, \bibinfo {author} {\bibfnamefont {P.}~\bibnamefont {Zoller}}, \ and\
  \bibinfo {author} {\bibfnamefont {H.}~\bibnamefont {Ritsch}},\ }\href
  {\doibase 10.1103/PhysRevA.45.4879} {\bibfield  {journal} {\bibinfo
  {journal} {Phys. Rev. A}\ }\textbf {\bibinfo {volume} {45}},\ \bibinfo
  {pages} {4879} (\bibinfo {year} {1992})}\BibitemShut {NoStop}%
\bibitem [{\citenamefont {{Gadway}}\ \emph {et~al.}(2012)\citenamefont
  {{Gadway}}, \citenamefont {{Pertot}}, \citenamefont {{Reeves}},\ and\
  \citenamefont {{Schneble}}}]{Gadway2011a}%
  \BibitemOpen
  \bibfield  {author} {\bibinfo {author} {\bibfnamefont {B.}~\bibnamefont
  {{Gadway}}}, \bibinfo {author} {\bibfnamefont {D.}~\bibnamefont {{Pertot}}},
  \bibinfo {author} {\bibfnamefont {J.}~\bibnamefont {{Reeves}}}, \ and\
  \bibinfo {author} {\bibfnamefont {D.}~\bibnamefont {{Schneble}}},\ }\href
  {http://dx.doi.org/10.1038/nphys2320} {\bibfield  {journal} {\bibinfo
  {journal} {Nature Physics}\ }\textbf {\bibinfo {volume} {8}},\ \bibinfo
  {pages} {544} (\bibinfo {year} {2012})}\BibitemShut {NoStop}%
\bibitem [{\citenamefont {Altman}\ \emph {et~al.}(2004)\citenamefont {Altman},
  \citenamefont {Demler},\ and\ \citenamefont {Lukin}}]{Altman2004a}%
  \BibitemOpen
  \bibfield  {author} {\bibinfo {author} {\bibfnamefont {E.}~\bibnamefont
  {Altman}}, \bibinfo {author} {\bibfnamefont {E.}~\bibnamefont {Demler}}, \
  and\ \bibinfo {author} {\bibfnamefont {M.~D.}\ \bibnamefont {Lukin}},\ }\href
  {\doibase 10.1103/PhysRevA.70.013603} {\bibfield  {journal} {\bibinfo
  {journal} {Phys. Rev. A}\ }\textbf {\bibinfo {volume} {70}},\ \bibinfo {eid}
  {013603} (\bibinfo {year} {2004})}\BibitemShut {NoStop}%
\bibitem [{\citenamefont {Di\'osi}\ and\ \citenamefont
  {Kiefer}(2000)}]{Kiefer2000a}%
  \BibitemOpen
  \bibfield  {author} {\bibinfo {author} {\bibfnamefont {L.}~\bibnamefont
  {Di\'osi}}\ and\ \bibinfo {author} {\bibfnamefont {C.}~\bibnamefont
  {Kiefer}},\ }\href {\doibase 10.1103/PhysRevLett.85.3552} {\bibfield
  {journal} {\bibinfo  {journal} {Phys. Rev. Lett.}\ }\textbf {\bibinfo
  {volume} {85}},\ \bibinfo {pages} {3552} (\bibinfo {year}
  {2000})}\BibitemShut {NoStop}%
\bibitem [{\citenamefont {Ollivier}\ \emph {et~al.}(2004)\citenamefont
  {Ollivier}, \citenamefont {Poulin},\ and\ \citenamefont
  {Zurek}}]{Zurek2004a}%
  \BibitemOpen
  \bibfield  {author} {\bibinfo {author} {\bibfnamefont {H.}~\bibnamefont
  {Ollivier}}, \bibinfo {author} {\bibfnamefont {D.}~\bibnamefont {Poulin}}, \
  and\ \bibinfo {author} {\bibfnamefont {W.~H.}\ \bibnamefont {Zurek}},\ }\href
  {\doibase 10.1103/PhysRevLett.93.220401} {\bibfield  {journal} {\bibinfo
  {journal} {Phys. Rev. Lett.}\ }\textbf {\bibinfo {volume} {93}},\ \bibinfo
  {pages} {220401} (\bibinfo {year} {2004})}\BibitemShut {NoStop}%
\bibitem [{\citenamefont {Blume-Kohout}\ and\ \citenamefont
  {Zurek}(2006)}]{Zurek2006a}%
  \BibitemOpen
  \bibfield  {author} {\bibinfo {author} {\bibfnamefont {R.}~\bibnamefont
  {Blume-Kohout}}\ and\ \bibinfo {author} {\bibfnamefont {W.~H.}\ \bibnamefont
  {Zurek}},\ }\href {\doibase 10.1103/PhysRevA.73.062310} {\bibfield  {journal}
  {\bibinfo  {journal} {Phys. Rev. A}\ }\textbf {\bibinfo {volume} {73}},\
  \bibinfo {pages} {062310} (\bibinfo {year} {2006})}\BibitemShut {NoStop}%
\bibitem [{\citenamefont {{Zurek}}(2007)}]{Zurek2007a}%
  \BibitemOpen
  \bibfield  {author} {\bibinfo {author} {\bibfnamefont {W.~H.}\ \bibnamefont
  {{Zurek}}},\ }\href@noop {} {\bibfield  {journal} {\bibinfo  {journal} {ArXiv
  e-prints}\ } (\bibinfo {year} {2007})},\ \Eprint
  {http://arxiv.org/abs/0707.2832} {arXiv:0707.2832 [quant-ph]} \BibitemShut
  {NoStop}%
\bibitem [{\citenamefont {Riedel}\ and\ \citenamefont
  {Zurek}(2010)}]{Zurek2010a}%
  \BibitemOpen
  \bibfield  {author} {\bibinfo {author} {\bibfnamefont {C.~J.}\ \bibnamefont
  {Riedel}}\ and\ \bibinfo {author} {\bibfnamefont {W.~H.}\ \bibnamefont
  {Zurek}},\ }\href {\doibase 10.1103/PhysRevLett.105.020404} {\bibfield
  {journal} {\bibinfo  {journal} {Phys. Rev. Lett.}\ }\textbf {\bibinfo
  {volume} {105}},\ \bibinfo {pages} {020404} (\bibinfo {year}
  {2010})}\BibitemShut {NoStop}%
\end{thebibliography}

%

\end{document}